| Title | Influence of surface stoichiometry and quantum confinement on the electronic structure of small diameter $In_xGa_{1-x}As$ nanowires |
|---|---|
| Author(s) | Razavi, Pedram; Greer, James C. |
| Original citation | Pedram Razavi, James C. Greer, "Influence of surface stoichiometry and quantum confinement on the electronic structure of small diameter InxGa1-xAs nanowires", Materials Chemistry and Physics, Volume 206, 2018, Pages 35-39, ISSN 0254-0584, doi: 10.1016/j.matchemphys.2017.12.006 |
| Type of publication | Accepted Manuscript (peer-reviewed)<br>To appear in: *Materials Chemistry and Physics* |
| Link to publisher's version | https://doi.org/10.1016/j.matchemphys.2017.12.006<br>Access to the full text of the published version may require a subscription |
| Copyright information | © 2018. This manuscript version is made available under the CC-BY-NC-ND 4.0 license http://creativecommons.org/licenses/by-nc-nd/4.0/ |

# Influence of surface stoichiometry and quantum confinement on the electronic structure of small diameter $In_xGa_{1-x}As$ nanowires

Pedram Razavi and James C. Greer

Tyndall National Institute, University College Cork, Lee Maltings, Dyke Parade, Cork, Ireland T12 R5CP

**Corresponding author email: pedram.razavi@tyndall.ie**

**Abstract**

Electronic structures for $In_xGa_{1-x}As$ nanowires with [100], [110], and [111] orientations and critical dimensions of approximately 2 nanometer are treated within the framework of density functional theory. Explicit band structures are calculated and properties relevant to nanoelectronic design are extracted including band gaps, effective masses, and density of states. The properties of these III-V nanowires are compared to silicon nanowires of comparable dimensions as a reference system. In nonpolar semiconductors, quantum confinement and surface chemistry are known to play a key role in the determination of nanowire electronic structure. $In_xGa_{1-x}As$ nanowires have in addition effects due to alloy stoichiometry on the cation sublattice and due to the polar nature of the cleaved nanowire surfaces. The impact of these additional factors on the electronic structure for these polar semiconductor nanowires is shown to be significant and necessary for accurate treatment of electronic structure properties.

# 1. Introduction

Transistor scaling results in a set of deleterious performance issues in general referred to as short-channel effects (SCEs). The onset of SCEs led to the requirement for various technology boosters to maintain continued transistor scaling while achieving enhanced performance [1]. Scaling has led to fin-like three dimensional field effect transistors, or FinFETs and most advanced device roadmaps are now including nanowire transistors for 4 nanometer technologies (although some care is required for defining critical dimensions for new technology "nodes"). Nanowires with gate-all-around architectures and with very small cross-sections provide excellent electrical characteristics with low standby power, and increase the potential for high-density integration. For nanowire diameters less than the Fermi wavelength of free charge carriers, quantum mechanical effects become pronounced. At these critical dimensions, quantum confinement leads to dramatically different electronic band structures with quite different band gaps and effective masses as compared to the same material in bulk form. As channel lengths decrease, carrier scattering is reduced and bulk properties such as charge carrier mobility must be abandoned or redefined for the quasi-ballistic regime.

Due to significantly higher bulk electron and hole mobilities, III-V materials such as $In_{0.53}Ga_{0.47}As$ and the group IV material germanium, respectively, have been suggested as a replacement for silicon as the channel material in advanced complimentary metal-oxide-semiconductor (CMOS) fabrication processes. Other III-V materials such as InAs NWs have also attracted considerable interests for their potential in high-performance applications in optoelectronics [2], as well as low-power logic applications [3]. Notwithstanding the technological challenges of replacing silicon in conventional semiconductor manufacturing, there is comparatively little known about the electronic structure of these replacement materials when patterned or grown in nanowires structures with diameters of the order of a few nanometer. Due to advances in bottom up as well as top down fabrication techniques, nanowires with diameters as small as 1 nm have been fabricated. To highlight recent work, we note the preparation of SiNWs with approximately 1 nm diameters for which the oxide sheath was removed and replaced with hydrogen termination; using scanning tunneling microscopy (STM) the band gap widening due to quantum confinement was reported [4]. Recently using a metalorganic chemical vapor deposition (MOCVD) via vapor-liquid-solid (VLS) growth method with gold nanoparticle catalysis, InAs NWs with approximately 2 nm diameter have been observed [5]. This study focuses on $In_xGa_{1-x}As$ nanowires with different crystallographic orientations and cross sections on the order of 5 square nanometer. Alloying on the cation sublattice is considered to be random as observed in the bulk. However, compound semiconductor growth is normally not under equilibrium conditions and the growth or fabrication conditions can lead to cation-rich surfaces or anion-rich surfaces. This leads to a new factor that impacts on the electronic structure of III-V compound semiconductor nanowires at small cross sections. The influence of either cation or anion atoms at the surface of the nanowires is explicitly considered in the calculations, and as will be shown, has a significant influence on the resulting electronic structure. There are previous reports of the electrical properties for InGaAs quantum wells and thin-body transistors [6,7,8]. However, studies of the electrical properties of $In_xGa_{1-x}As$ nanowires at dimensions of relevance for gate-all-around CMOS technologies with critical dimensions of 4 nanometer and less are not available. The critical dimensions (cross sectional area) considered are chosen to extract the interaction between quantum confinement and surface effects. The interplay between these effects is relevant for modern nanowire designs, particularly for nanoelectronics and sensing applications. We note that leading nanoelectronics manufacturers are currently developing devices for the 7 nm and 5 nm nodes [9,10,11]. The challenge for the research community is to explore the surface chemistry and confinement physics for nanowires with critical dimensions approaching atomic scale limits.

2. **Methods**

Figure 1 shows the relaxed structures for three crystallographic orientations of GaAs nanowires considered in this study; these structures form the basis for the generation of the nanowires with differing alloy composition. All atomistic visualizations are rendered using VESTA [12].

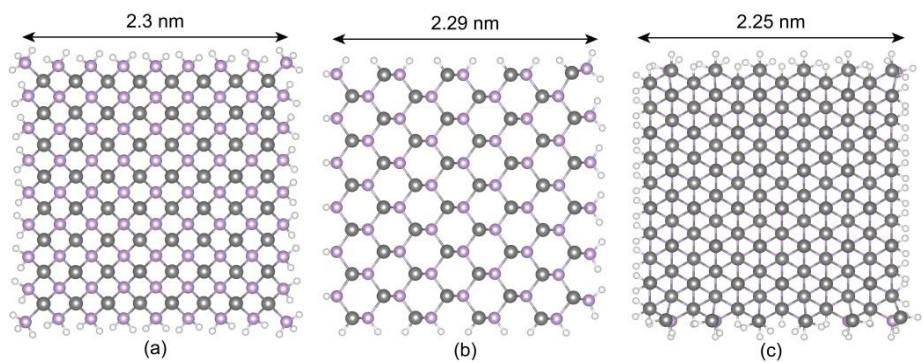

*Fig.1. Cross-section of* GaAs *NWs for (a)* [100]*, (b)* [110]*, and (c)* [111] *wire orientations composed of* As *(purple),* Ga *(grey) and pseudo-*H *(white) atoms. Note that pseudo-hydrogens of appropriate charge are chosen for each surface atom to provide a defect free surface.*

The electronic band structures of small cross section, approximately square, $In_XGa_{1-X}As$ nanowires are investigated for different alloy compositions and for different crystallographic orientations using density functional theory. The impact of different surface character (*i.e.* cation or anion) is also considered. Silicon nanowires of similar cross sections and with the same orientations as for the III-V nanowires are also calculated for comparison. Using the Kohn-Sham eigenvalues to approximate band structures, the effect of band folding due to quantum confinement is determined. Band folding results in significantly different electronic structures in the nanowires compared to their bulk form as is reflected in large changes to the band gap energies and effective masses at the conduction and valence band edges. Knowledge of these physical parameters that can be directly extracted from the electronic band structures is key to accurate calculations of charge transport through nanoscale transistors [7]. Electron effective masses govern charge transport and as a first approximation give an indication of the change in electron mobility as critical dimensions are reduced. The direct and indirect nature of a band gap is crucial for engineering of photonic devices; the magnitude of the band gap controls the thermal operation of a semiconductor in a transistor, and as well controls and/or limits source-to-drain tunneling.

A representative structure is selected at each nanowire alloy composition from a large set of geometries generated by randomly occupying each site of the cation sublattice with either an In or Ga atom with probability proportional to the alloy stoichiometry. The nearest neighbor distributions about each of the arsenic atoms in each nanowire generated in this manner is calculated leading to a distribution of In and Ga nearest neighbors. Specific nanowire configurations with nearest neighbor averages matching the positions of the peaks in the ensemble nearest neighbor distribution are chosen to represent the set of alloy configurations accessible at a given stoichiometry.

All calculations are performed using the Vienna *ab initio* simulation package (VASP) [13,14,15] using the Pedrew-Burk-Ernzerhof (PBE) exchange-correlation functional [16] and the projected augmented wave method [17,18]. A plane wave cut-off energy of 460 eV is chosen. A *k*-point grid of 11 x 1 x 1 is used to converge the electronic energy. The lattice parameters for all nanowires are relaxed and atomic positions are optimized to a force tolerance of less than 10 meV / Å per atom. A primary goal of the calculations is to understand the impact of the cation sublattice stoichiometry on electronic structure and physical properties as extracted from the band structures. The pseudo-hydrogen approach to surface terminations as described in, for example, ref. [19] is used. Pseudo-hydrogen atoms with fractional charges are widely used to chemically saturate surface dangling bonds in III-V materials to obtain an ideal passivation with no defect states in the nanowire band gap. Electrical properties of $In_XGa_{1-X}As$ nanowires are investigated by determining the band gaps and effective masses from the electronic structure calculations. The electron effective masses are calculated using an approximation of the second derivative $\partial^2 E/\partial k^2$ at the conduction band minima using a 5 point stencil method [20].

The band gap energy is an important factor in design of tunneling devices [21] and also has a significant impact on the ON-OFF current ratio in metal-oxide field effect transistors (MOSFETs) [22,23,24]. The effective masses can be considered in a first approximation as indicating the relative electron mobility for nanowires of different crystallographic orientation with a smaller effective mass suggesting higher electron mobility – assuming a similar order of magnitude in various electron relaxation processes between the various III-V nanowires examined. On the other hand, a small effective mass is concomitant with a lower density of states (DoS) leading to a lower inversion charge density [3] for device applications. Lower inversion charge density and phenomena such as band-to-band tunneling (BTBT) can severely limit device performance for transistors operating in a ballistic regime [22,25]. In addition to tabulating the conduction band effective masses, the calculated DoS is also reported for different nanowires to allow for comparison.

## 3. Results and Discussion

Fig. 2 shows the effect of alloy composition on the electronic band structure of the [100]-oriented $In_xGa_{1-x}As$ nanowires and the DoS for gallium stoichiometries of 1-x = 0, 0.5, and 1. Within fig. 2 a significantly lower DoS at the conduction band minima of the III-V nanowires relative to the silicon nanowire reference is found due to the much lower effective masses for the III-V nanowires, and as well due to a much larger energy separation between the two lowest energy conduction bands at the Γ-point in the Brillouin zone. As a result the DoS near the conduction band edges are significantly less in the $In_xGa_{1-x}As$ nanowires relative to the comparable silicon NW as compared in fig. 2b. Consistent with this observation is that the calculated effective mass of the [100]-oriented silicon nanowire is about 4 to 6 times larger than for the $In_xGa_{1-x}As$ nanowires indicating a much higher electron mobility for the $In_xGa_{1-x}As$ nanowires.

To contrast the electronic structures of the InAs and GaAs NWs with the hydrogen passivated Si nanowire of similar cross section, the electronic band structures for GaAs, InAs and Si NWs are displayed in fig. 3 for [100], [110], and [111] nanowire orientations. The extracted band gap energies and effective masses at the Γ-point for the various $In_xGa_{1-x}As$ alloy compositions with 1-x varied from 0 to 1 in increments of 0.25 are displayed in Fig. 4. Fig. 4(a) illustrates that [110]-oriented nanowires, which have anion-rich surfaces due to the two polar facets as can be seen in fig. 1(b), have smaller band gaps and effective masses whereas the [100]- and [111]-oriented nanowires have similar band gaps to one another. On the other hand, [111]-oriented wire are predicted to have smaller effective masses at the conduction band edge than the [100] nanowires. For comparison, the [100]-oriented and [110]-oriented silicon nanowires have direct band gaps of 1.17eV and 0.95 eV, respectively. However, the [111]-oriented silicon nanowire has an indirect band gap which is consistent with previously reported silicon nanowire band structures [26]. Note that the calculated effective mass of [100]- and [110]-oriented silicon nanowire at gamma are $0.35m_e$ and $0.13m_e$, respectively.

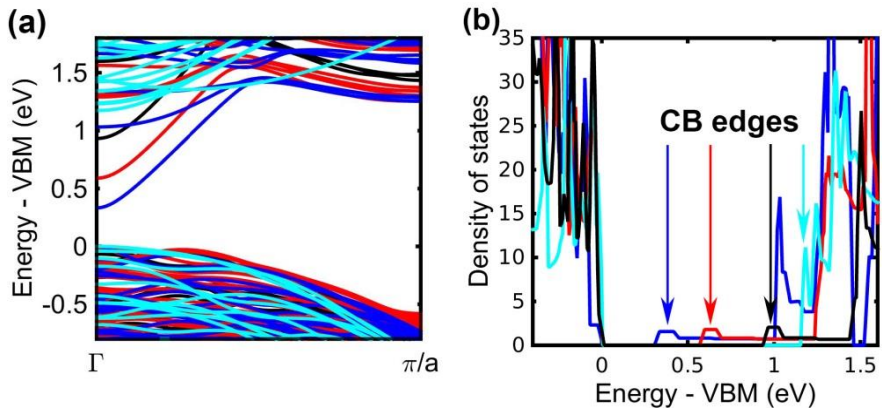

*Figure 2. Effect of alloy composition for [100]-oriented $In_xGa_{1-x}As$ on the (a) band structure, and (b) the density of states for gallium stoichiometries of 1-x = 0 (blue) or InAs, 0.5 (red), and 1 (black) or GaAs as compared with a similar cross section [100]-oriented silicon NW (cyan).*

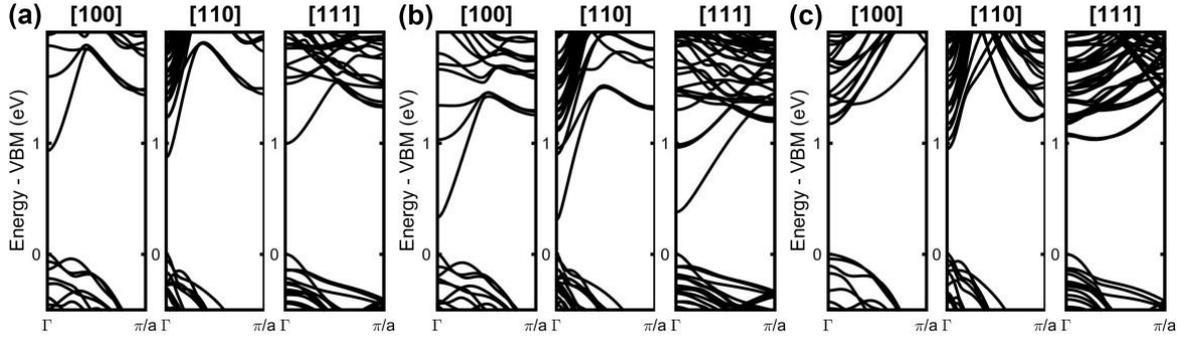

*Figure 3. Electronic band structures for (a) GaAs, (b) InAs and (c) Si nanowires with crystallographic orientations as labeled.*

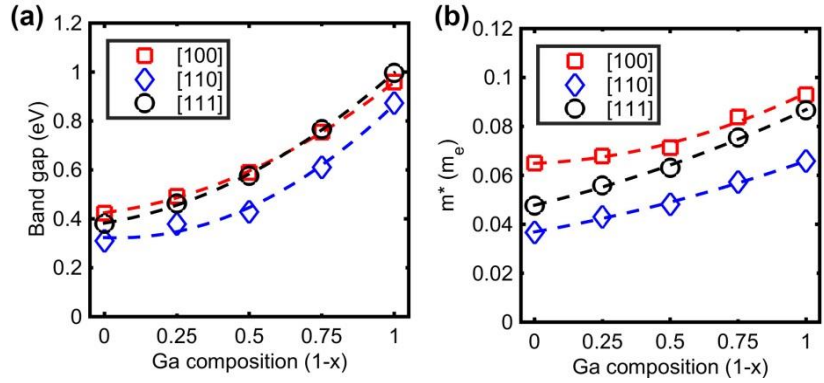

*Figure 4. Comparison of (a) band gaps and (b) effective masses for $In_xGa_{1-x}As$ NWs with different gallium compositions 1-x and crystallographic orientations as labeled in the figure legends.*

A factor that arises in III-V semiconductor nanowires not featuring in nanowires, for example, composed of group IV nor alloys of group IV atoms, is the surface of the III-V nanowires can be anion-rich, cation-rich or intermediate between these limits. The effect of the surface character on the electronic band structures is studied in [100]-oriented InAs and GaAs nanowires with a primarily anion-rich and cation-rich semiconductor surfaces in the nanowires, again pseudo-hydrogen termination has been used to eliminate surface states. The resulting effective masses and energy band gaps for this set of nanowires are listed in Table 1.

It can be seen that in both compositions InAs and GaAs, the anion-terminated surfaces result in smaller band gaps and lower effective masses relative to the nanowires constructed with a cation-rich semiconductor surface. In contrast to bulk GaAs compared to bulk silicon, at the dimensions considered the [100]-oriented GaAs nanowire can display a smaller band gap than the comparable silicon nanowire which is found to be 1.17 eV with hydrogen termination although caution is needed due to the deficiencies of approximate DFT in predicting band gap energies. The effective masses in the III-V nanowires, as in the bulk, remain significantly lower than the effective mass at the conduction band edge for the comparable cross section [100]- oriented silicon nanowire of $m^*/m_e=0.35$. The fact that surface termination can significantly affect band gap energies is well-known for non-polar semiconductors [27] and semimetals [28]. Here it is highlighted that the breaking of periodicity in a polar nanowire can lead to either anion- or cation-rich surfaces that, in addition to the polarity induced by surface terminating species, can significantly influence nanowire electronic structure properties.

*Table.1 Band gap ($E_g$) and effective mass ($m^*$) of InAs and GaAs NWs at the $\Gamma$-point with surfaces that are either anion- or cation-rich. Band gap energies in electron volts and effective masses are given relative to the free electron rest mass. For comparison, the band gap and effective mass of the silicon nanowire with hydrogen passivation for approximately the same cross section is calculated to be 0.35 $m_e$ and 1.17 eV, respectively.*

|   |   | [100]-oriented NWs | |
|---|---|---|---|
|   |   | Surface anions | Surface cations |
| InAs NW | $m^* / m_e$ | 0.065 | 0.166 |
|         | $E_g$ / eV  | 0.423 | 0.655 |
| GaAs NW | $m^* / m_e$ | 0.092 | 0.166 |
|         | $E_g$ / eV  | 0.958 | 1.152 |

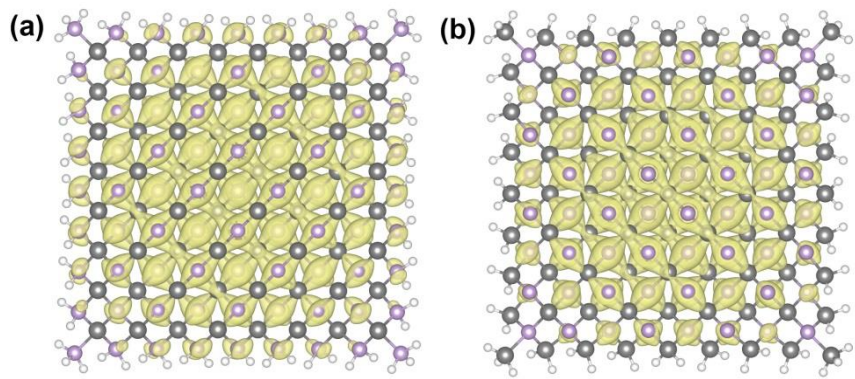

*Figure 5. Charge density for the electronic states at the conduction band minima for the (a) anion-terminated, and (b) cation-terminated [100]-oriented GaAs nanowires. Atoms are shown as: As (purple), Ga (grey) and pseudo-H (white).*

To explore the variation of the electronic charge density with differing surface character, the electronic state at the conduction band minima for [100]-oriented GaAs nanowires is shown in fig. 5. It is well known for GaAs that of the (100), (110) and (111) surfaces only the (110) plane can be constructed as non-polar. The (100) and (111) surfaces can be constructed as either anion- or cation-rich. In fig. 5, the surfaces are either (a) anion-rich or (b) cation-rich {100} surfaces. As seen in fig. 5, there is considerable charge localized onto As atoms throughout the nanowire cross section for the anion-rich layers at the surfaces. This is contrasted to the case in fig. 5(b) where the surface is cation-rich and the charge does not localize at the surface layers. Hence the confinement within the nanowire with cation-rich surfaces is larger, and a higher conduction state energy is expected relative to the anion-rich nanowire surfaces. This result is consistent with the larger band gap energies for the cation-rich layers at the surface compared to of the anion-terminated nanowires given in table 1. The different electronegativity of the differing surface layers leads to differing surface dipole potentials. For the III-V nanowires variation in the surface layer composition leads to band gap shifts of smaller magnitude than that found for differing surface terminations in non-polar nanowires [27,28], but is nonetheless a significant influence on the electronic structure.

The effect of introducing different surface layers into [110]-oriented $In_{1-x}Ga_xAs$ nanowires is considered. Two polar surfaces and two non-polar surfaces are obtained. The two polar surfaces are chosen to either be anion- and cation-rich and are labeled structure 1 (Str 1) and structure 2 (Str 2), respectively. The cross-section for the [110]-oriented GaAs nanowires with the different surface layers is shown in fig. 6(a) and the effective masses obtained at the conduction band edge at the Γ-point are compared in fig. 6(b). The overall potential due to the two polar surfaces is smaller than in compared to the case of the [100]-nanowires with four polar surfaces considered in fig. 5. Nonetheless the influence of the polar surfaces is still significant. The difference in effective mass along the nanowire axis caused by having either two anion-rich or two cation-rich surfaces at a given alloy composition is the same order of magnitude as the overall change due to differing alloy composition. The nanowire with the anion-rich surface layers provides the lower effective mass between the two structures.

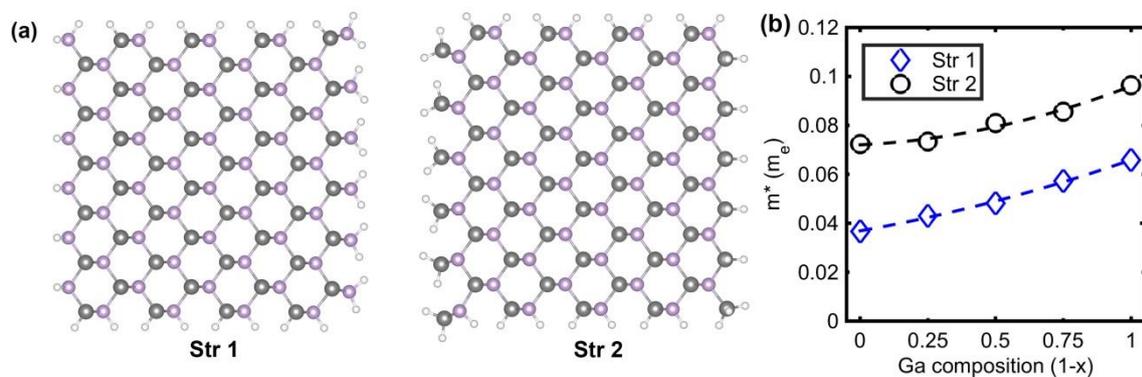

*Figure 6. (a) Cross-section of [110]-oriented GaAs with two different structures. Structure 1 (Str 1): anion-rich on two polar surfaces and with two non-polar surfaces. Structure 2 (Str 2): cation-rich on two polar surfaces and with two non-polar surfaces. (b) Calculated effective masses for [110]-oriented $In_xGa_{1-x}As$ as a function of alloy composition. The atoms are colored as* As (purple)*,* Ga (grey) *and pseudo-*H (white)*.*

4. Conclusions

The electronic structures of $In_xGa_{1-x}As$ nanowires with cross sections of approximately 5 square nanometer have been computed and band gaps, effective masses, and the density of states have been extracted from the energy bands. The alloy stoichiometry on the cation sublattice has been varied and the [100], [110] and [111] nanowire orientations have been studied. Even at these extremely small critical dimensions, smaller effective masses for these III-V nanowires are observed relative to silicon nanowires of comparable dimensions. As is the case also for the bulk materials, the III-V nanowires show a low density of states relative to the comparable silicon nanowire. Hence design constraints related to the use of III-V materials in macroscopic field effect transistors are anticipated to persist even for these scaled dimensions.

In addition to quantum confinement and surface chemistry effects, the compound semiconductor nanowires exhibit additional influences on their electronic structure due to alloying on the cation sublattice and due to different surface layer compositions leading to either anion- or cation-rich polar surfaces. The latter two factors are not present for nonpolar semiconductor nanowires but are demonstrated here to have a significant influence on the electronic properties of ultra-thin nanowires. For design of nanowire transistors and sensors, these additional effects must be explicitly accounted for accurate treatment of the electronic band structures as required for accurate device design.

**Acknowledgments**
This work was supported by the European Union project DEEPEN funded under NMR-2013-1.4-1 grant agreement number 604416 and partially funded by the Science Foundation Ireland SMALL project 13/IA/1956. We also wish to acknowledge the SFI/HEA Irish Centre for High-End Computing (ICHEC) for the provision of computational facilities.